\documentclass{aa}

\usepackage{rotate} 
\input rotate.tex
\input psfig
  
\def\arcmin{\hbox{$^\prime$}}

\def\fdg{\hbox{$.\!\!^\circ$}}

\def \hi {H\,{\sc i~}} 
 
\def\kms{km\,s$^{-1}$}

\begin{document} 
\thesaurus{11(09.11.1; 11.04.2; 11.09.1: Sgr Dwarf Spheroidal; 
11.09.2; 11.09.4)} 
\title{HI Observations Towards the Sagittarius Dwarf Spheroidal Galaxy}
\titlerunning{Sagittarius Dwarf Galaxy}
\author{W. B. Burton\inst{1} and Felix J. Lockman \inst{2}} 
\institute{Sterrewacht Leiden, P.O. Box 9513, 2300 RA Leiden, The Netherlands,
\\ burton@strw.leidenuniv.nl 
\and National Radio Astronomy Observatory, P.O. Box 2, Green Bank WV 24944,
USA, \\ jlockman@nrao.edu} 

\date{Received mmdd, 1999; accepted mmdd, 1999} 
\offprints{W.B. Burton or F.J. Lockman} 
\maketitle 

\begin{abstract} 
  
  We have measured the $\lambda$21 cm line of Galactic \hi over
more than 50 square degrees in the 
direction of the Sagittarius dwarf spheroidal galaxy.  
The data show no evidence
of \hi associated with the dwarf spheroidal which might be consider analogous 
to the Magellanic Stream as it is associated in both position and velocity 
with the Large Magellanic Cloud.  Nor do the \hi data show evidence for any
disturbance in the Milky Way disk gas that can be unambiguously assigned 
to interaction with the dwarf galaxy. 
The data shown here limit the  \hi mass at the velocity of the Sagittarius 
dwarf to  $<7000$ M$_\odot$ over some
18 square degrees between Galactic latitudes $-13^\circ$ and
$-18\fdg5$.

\end{abstract} 
\section{Introduction} 
\label{intro}

The Sagittarius dwarf spheroidal 
galaxy appears as a stream of stars covering
several hundreds of square degrees around 
$l=5^\circ$, $b=-15^\circ$.  It is elongated perpendicular
to the Galactic plane  and apparently is in the process
of plunging through the Milky Way some 25 kpc from the Sun 
and 17 kpc from the Galactic center (Ibata et al.
1994, 1997; Mateo et al. 1998).
This encounter is probably  not the first for this system.  
It seems reasonable to search for effects which such an encounter
might plausibly produce on the Milky Way, including distension of the  \hi
layer in the direction of the dwarf galaxy, and, depending on the dwarf's 
mass, creation or enhancement of a  Galactic warp 
(Lin \cite{lin96}, Ibata \& Razoumov \cite{ibat98}).  
Previous $\lambda21$~cm measurements have detected no \hi
at three positions near the center
of the Sgr dwarf (Koribalski et al. \cite{kori94}), 
making this system typical of nearby dwarf spheroidals, although 
the dSph galaxy Sculptor is an interesting exception (Carignan et al. 
\cite{cari98}, Carignan \cite{cari99}).

We have measured the  $\lambda$21~cm emission line of neutral hydrogen
over a large area which 
includes the central region of the Sgr dwarf
 in a search for 
neutral gas associated with or entrained by the dwarf galaxy, 
for a perturbation in Galactic gas 
resulting from the first phases of an encounter,
or for evidence for other phenomena such as a tidal tail
or an analogue to the Magallenic Stream as it is associated with the Large
Magellanic Cloud.

\section{Observations} 
\label{sec:obs}

Galactic \hi spectra were taken toward the Sagittarius
 dwarf galaxy using the 140 Foot
Telescope of the NRAO in Green Bank, WV, which has an angular
resolution of $21\arcmin$ at the 21cm wavelength
of the \hi line. 
Each spectrum covers about 1000 km~s$^{-1}$ centered at a velocity of
$+50$ km~s$^{-1}$ with respect to the LSR.  The channel spacing is 
2.0 km~s$^{-1}$.  The data were obtained by switching against a
reference frequency $+5$ MHz above the signal band.  
  The temperature scale was calibrated
using laboratory measurements of a noise diode.
Spot checks show that the spectra
do not deviate by more than 10\% from the calibrated intensity scale of 
 the Leiden/Dwingeloo \hi survey of Hartmann \& Burton (1997). 
 The system temperature at zenith was about 20 K;
a typical rms noise level in a spectrum is about 
30 mK.  This noise is 
equivalent to a column density uncertainty 
$\sigma(N_{\rm {HI}})=1.8 \times 10^{17}$ cm$^{-2}$ over a 20 
km~s$^{-1}$ band.
At the 25 kpc distance of the Sagittarius Dwarf Spheroidal, this 
noise level allows a $3\sigma$ detection of 200 M$_\odot$ of \hi 
 in a single 140 Foot  pointing.

Spectra were taken every $0\fdg5$ in Galactic longitude and latitude
over the region bounded by $4\fdg0 \leq l \leq 8\fdg0$ 
and $-4\fdg0 \leq b \leq -15\fdg5$, 
with an extension to $b=-18\fdg5$ over 
$4\fdg0 \leq l \leq 6\fdg5$.  In all, 
more than 250 \hi spectra were measured.

We also examined \hi spectra that were available over a more extended region,
albeit taken at lower sensitivity, from the Leiden/Dwingeloo survey
which covers the sky north of declination $-30^\circ$ at half--degree
spacing, and from the observations of Liszt
 \& Burton (1980), which cover the area within about $10^\circ$ of
the Galactic center, also at half--degree intervals.

\section{Expected Location of  HI Emission}

\begin{figure}[t]
\psfig{file=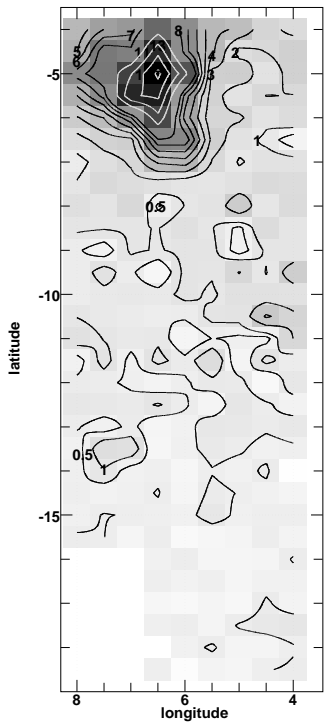,width=14cm,angle=270}
\vfill
\caption{Contours and grey scale of neutral hydrogen integrated over 
+130 \kms to +170 \kms\, over the region of our observations. 
The \hi column descreases smoothly from the Galactic plane, and
there is no indication of significant \hi associated with the
Sagittarius dwarf spheroidal galaxy in these data. The contours are
labelled in units of K \kms, each equivalent to $1.8 \times 10^{18}$ 
cm$^{-2}$.}
\end{figure}

\begin{figure}[t]
\psfig{file=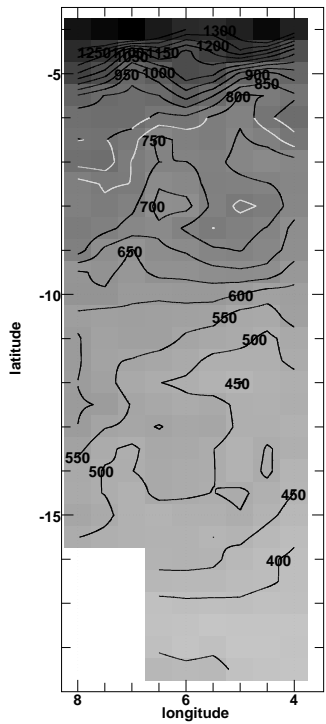,width=14cm,angle=270}
\vfill
\caption{Contours and grey scale of neutral hydrogen integrated over 
the range $-400$ \kms\, to $+500$ \kms. 
The \hi column descreases smoothly from the Galactic plane. 
There is no indication of significant \hi associated with the
Sagittarius dwarf spheroidal galaxy over this extended velocity range. The 
contours are labelled in units of K \kms, 
each equivalent to $1.8 \times 10^{18}$ 
cm$^{-2}$.}
\end{figure}

\begin{figure*}[t]
\psfig{file=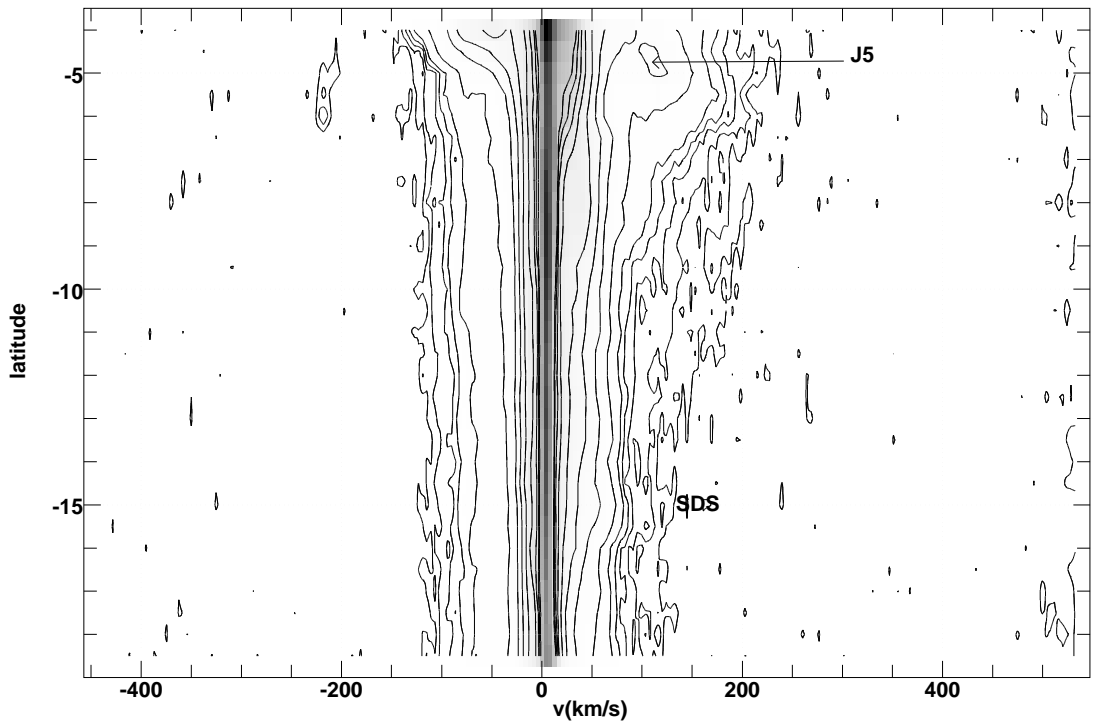,width=20cm,angle=-90}
\vfill
\vspace{-5cm}
\caption{Latitude--velocity diagram showing contours of HI brightness
temperature averaged at each latitude over longitudes $4^\circ$ to 
$8^\circ$.  The LSR velocity and position of the brightest stellar component 
of the Sagittarius Dwarf Spheroidal galaxy is marked by the initials SDS.  
The width spanned by the initials is plus and minus twice the velocity
dispersion of the stellar component; the height of the initials corresponds  
approximately to the angular resolution of the 140 Foot Telescope.
 The \hi cloud marked J5 (Cohen 1974) is
likely associated with gas patterns associated with the Galactic bulge region
(Liszt \& Burton 1980). Contours are drawn at brightness--temperature 
emission levels of 0.015, 0.03, 0.05, 0.07, 0.1, 0.2, 0.5, 1.0, 2.0, 
4.0, 6.0, 8.0, and 10.0 K.}
\end{figure*}

\begin{figure*}[t]
\psfig{file=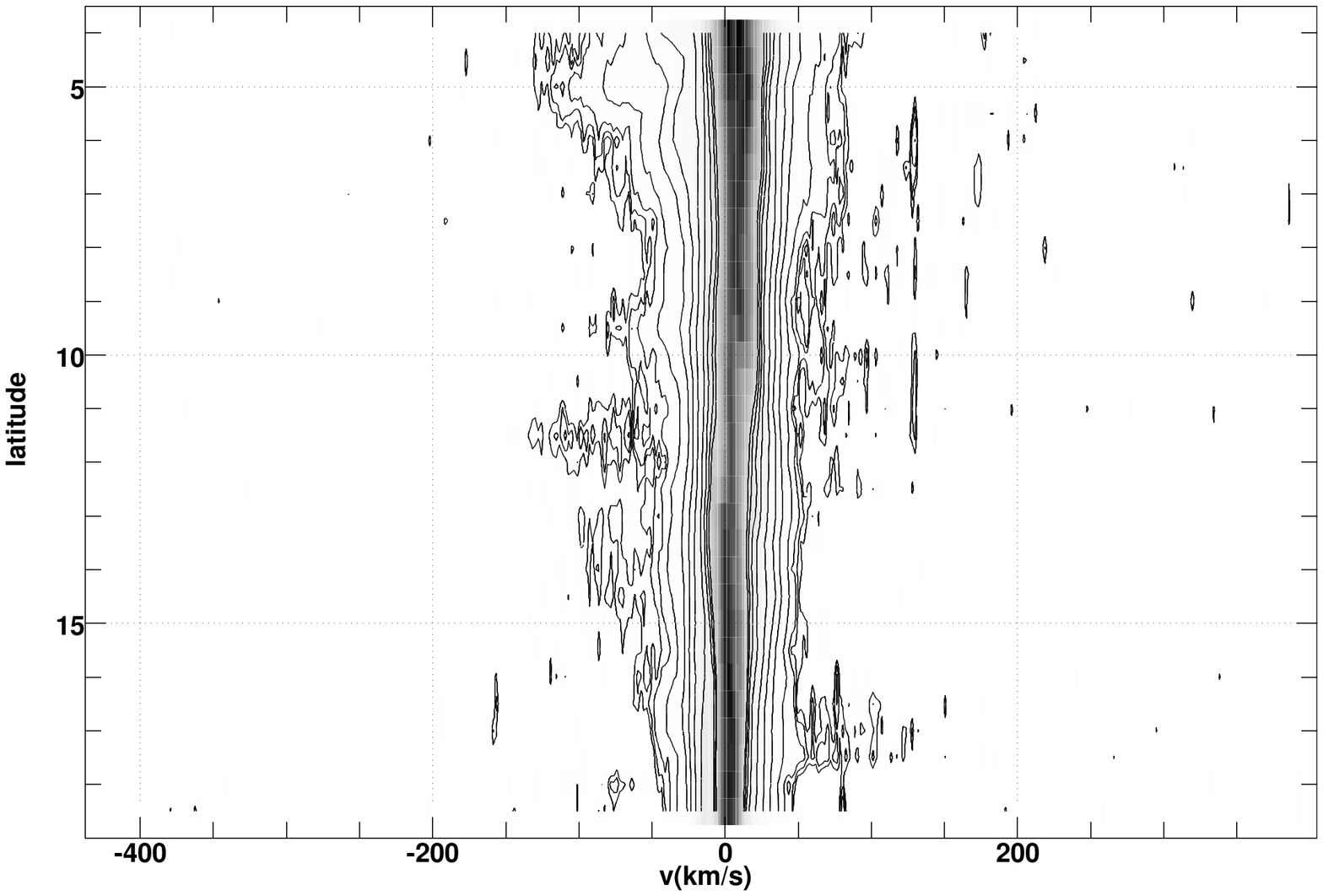,width=20cm,angle=-90}
\vfill
\vspace{-5cm}
\caption{Latitude--velocity diagram showing contours of HI brightness
temperature averaged at constant latitudes over longitudes $4^\circ$ to 
$8^\circ$ for regions {\it above} the Galactic plane.  The data are from the 
Leiden/Dwingeloo survey.  For easier comparison
with Figure 3 the latitude scale has been inverted. 
Contours are the same as for Figure 3, but the lowest
two contours have been omitted because the noise level of the data is higher 
here.}
\end{figure*}

The brightest parts  of the Sagittarius dwarf spheroidal
galaxy are near  $l = +5^{\circ}, b = -15^\circ$;
 stars associated with this system have been found from 
latitudes $-4^\circ$ to perhaps $-40^\circ$, and its true
extent is still uncertain (Alard \cite{alar96}, Alcock et al.
 \cite{alco97}, Mateo et al. \cite{mate98}).
The mean stellar velocity over the central region  is 
within a few km~s$^{-1}$ of +170 km~s$^{-1}$ with respect to 
the Galactic Standard of Rest, which, at the center of the dwarf,
corresponds to +150 km~s$^{-1}$ 
with respect to the Local Standard of Rest (Ibata et al. 1997).  
At such a low Galactic longitude, conventional Galactic rotation 
is almost entirely perpendicular to the line of sight, so unperturbed 
Milky Way disk gas at the 
location of the dwarf galaxy would have an LSR velocity of 
order $-10$ km~s$^{-1}$.  The longitude of the Sgr dwarf spheroidal passes,
however, close to the Galactic nucleus where 
 gas in the bulge regions displays both 
high positive-- and negative--velocity motions  
within about $6^\circ$ of the Galactic plane (e.g. Liszt and 
Burton \cite{lisz80}). 
Any \hi that is related to the Sagittarius dwarf would be expected to be
either at the velocity of the dwarf (in the Galactic Center Frame) 
or to be spatially extended in relation to the dwarf.  The former
possibility can be tested simply, because there is not expected to be Galactic 
\hi at the velocity of the dwarf; the later possibility requires comparison of
the Galactic \hi toward the dwarf with the situation in directions away 
from it.

\section{Results}
\label{sec:results}

Figure 1 shows the \hi intensity 
integrated over +130 to +170 \kms\, (LSR); this velocity range 
extends $\pm2\sigma$ about the mean stellar velocity 
of the Sagittarius dwarf spheroidal (Ibata et al. \cite{ibat97}).  
The amount of Galactic \hi decreases smoothly with latitude
over this area; there is no evidence for a concentration at the
latitude of the central part of the 
Sagittarius dwarf within this velocity range.  
A comparable view of the \hi
is given in Figure 2, which shows contours of
\hi intensity integrated over all observed velocities.  Again,
there is no significant concentration of \hi at the location of
the Sagittarius dwarf spheroidal within that extended velocity range.

Figure 3 shows a latitude--velocity map of the HI 
made after spectra at constant latitude have been 
averaged over the longitude range of our observations.  The 
centroid of the Sagittarius dwarf spheroidal galaxy is
marked by the letters `SDS'.

For comparison, Figure 4
shows a similar display of data for directions {\it above} the Galactic
plane from the Leiden/Dwingeloo survey of Hartmann \& Burton (1997). Note 
that the positive--latitude figure is plotted with the latitude scale inverted 
to facilitate comparison with the 
data toward the dwarf galaxy.
  
These figures show again that there is no 
bright \hi feature  at the
location and velocity of the dwarf galaxy.  Moreover, 
at the latitude of the Sagittarius dwarf spheroidal the 
appearance of the \hi contours at all velocities 
is quite similar above and below
the Galactic plane, showing no indication of a disturbance in
the Milky Was disk gas stemming from the proximity of the dwarf.  

The feature labeled J5 in Figure 3 is a cloud found by
Cohen (\cite{cohe75}) that has previously been interpreted
as one of the observational vagaries stemming from the projection of 
kinematics associated with the barred potential of the inner few kpc of the 
Galaxy (see Burton \& Liszt 1978, Liszt \& Burton 1980, Burton \& Liszt 1983).
J5 has a velocity with respect to the Galactic center of
$+140$ \kms, while at its latitude the Sagittarius dwarf
spheroidal has a velocity of about $+170$ \kms\, in the same
coordinate system.  Because of the velocity difference and lack of any
evident spatial coincidence, and
because feature J5 fits naturally into models of the Galactic
barred potential, we believe that an association of this 
\hi feature
with the Sagittarius dwarf spheroidal galaxy is unlikely.
The Liszt \& Burton model of the Galactic core also  
accounts for the small \hi feature at $b \sim 
+5^\circ$, $v \sim -75$ \kms\, seen in Figure 4 at the location and velocity
mirror--symmetric to the J5 feature.  Feature
J5 and its mirror--symmetric counterpart are a single pair of more than
the dozen apparently anomalous paired \hi features lying within the Galactic 
core 
discussed by Burton \& Liszt (1978).  All important geometrical aspects of 
the barlike model of the inner--Galaxy gas distribution proposed by Liszt
\& Burton (1980) whose consequences subsume the individual anomalous \hi
features, have subsequently been confirmed by the direct evidence for a
stellar bar in the Milky Way shown by the near--IR surface photometry (see
review by Fux 1999 and references there, in particular Blitz \& Spergel 1991,
Dwek et al. 1995, and Binney et al. 1997).  The larger context into 
which J5 and its mirror--symmetric counterpart fit in a natural way makes 
a suggested association with the Sagittarius dwarf particularly unconvincing.

There is an additional HI cloud visible in the data 
at ($l,b,v)=+8^\circ, -4^\circ, -212$
\kms.  This is a compact high--velocity cloud (see Braun \& Burton 1999) 
discovered by Shane (see 
Saraber \& Shane 1974), with no evident connection either to the Milky Way 
or to the Sagittarius dwarf.

We can place a  limit on the \hi mass at the velocity of the Sagittarius
dwarf spheroidal galaxy of $\sim200$ M$_\odot$ ($3\sigma$) in a single
beam of the 140 Foot Telescope, assuming that the \hi would have
the velocity centroid and velocity dispersion ($\sim 10$ \kms) 
observed in the stellar component (Ibata et al. \cite{ibat97}).  
With the same assumpton, a
 limit on the total mass of HI associated with the system in
the region covered by the current observations 
was derived by averaging 
all \hi spectra with $b \leq -13^\circ$ and 
$4\fdg0~\leq l \leq~8\fdg0$.  
The average \hi spectrum thus obtained shows some emission
above the noise level over the velocity range $+140 < v < +160$ \kms,
in an amount which would be equivalent to 
 7000 M$_\odot$ of \hi at the 25 kpc distance of the
Sagittarius dwarf spheroidal, but Figure 3 shows that 
most of this emission is coming from the lowest latitudes,
and it is neither centered on, nor increasing in intensity toward,
the position of the Sagittarius dwarf
spheroidal galaxy.  Indeed, the amount of \hi in the velocity
interval is doubled if the average is taken 
over an area extending to $b=-11^\circ$
rather than to $-13^\circ$.  We conclude that the rather small
amount of emission detected in the large-area average is
most likely contributed by gas in the Galactic core (possibly via
the antenna sidelobes) rather
than by gas associated with the Sagittarius dwarf spheroidal galaxy.

\section{Discussion}

There is no bright \hi associated with the Sagittarius dwarf 
spheroidal galaxy, either at the velocity of the galaxy itself,
or signaling a perturbation to the Milky Way.  
 There is no indication of a train of gas analogous 
to the Magellanic Stream (see Mathewson et al. 1974) which is focussed at 
the position and velocity of the Large Magellanic Cloud. 
The absence of \hi in the
dwarf spheriodal is consistant with earlier, more limited,
 \hi measurement 
(Koribalski et al \cite{kori94}) and with 
observations of similar systems.
In general, the dwarf spheroidal companions of the Galaxy and
of M31 do not have a detectable amount of HI.  The one exception
is the Sculptor dwarf galaxy, which has $>3 \times 10^4$ M$_\odot$ 
of \hi located in two clouds each about 500 pc from the center of
that galaxy (Carignan et al \cite{cari98}, Carignan \cite{cari99}).
These clouds have peak column densities in excess of
a few times $10^{19} $ cm$^{-2}$,  an order of magnitude greater than any
feature in our data at the position and velocity 
of the Sagittarius dwarf spheroidal galaxy (Figure 1).

The brightest part of the Sagittarius dwarf spheriodal galaxy 
is now about 5 kpc below the plane of the Milky Way, and
some stars associated with it may be within 2 kpc of the disk.  It appears
that it will strike the Milky Way at a location near the 
line of nodes of the Galactic warp, where the disk \hi is expected to be 
distributed approximately symmetrically about $z=0$.  The
\hi at a galactocentric distance of 17 kpc has, on average, a 
scale-height of about 400 pc (Burton \cite{burt92}).  The
main part of the Sagittarius dwarf is thus unlikely to be encountering
the Milky Way gaseous disk gas as of yet,
consistent with our finding 
 that the Sagittarius dwarf has not yet affected the Milky Way \hi 
in any significant way.

\begin{acknowledgements} 
 The National Radio Astronomy Observatory is operated by Associated
Universities, Inc., under a cooperature agreement 
  with the U.\,S. National Science Foundation.
\end{acknowledgements} 

{}

\end{document}